\documentclass[12pt]{article}
\usepackage{epsf,amsmath,graphicx}
\newcommand{\cd}{\makebox[0.08cm]{$\cdot$}}

\title{The Centauro events as a result of induced pions emission}
{\author{V.A. Karmanov$^a$\thanks{e-mail: karmanov@sci.lebedev.ru,
karmanov@isn.in2p3.fr}\quad and
A.E. Kudryavtsev$^b$\thanks{e-mail:kudryavt@heron.itep.ru} \\
{\small \em $^a$Lebedev Physical Institute, Leninsky Prospekt 53, 119991
Moscow, Russia}
\\{\small \em  $^b$Institute of Theoretical and Experimental Physics,}\\
{\small \em B. Cheremushkinskaya 25, 117259 Moscow, Russia}}}
\begin{document}
\maketitle
\bibliographystyle{unsrt}

\centerline{\Large \bf Preface} \bigskip

Like in the laser beam, where the Bose-Einstein statistics of photons results
in  its high momentum coherence,  in the production of large number of pions
this statistics  may lead to strongly enhanced emission of pions with the
same charge.  Because of that, in production of a hundred pions, the events
without neutral pions or, on the contrary, without the charged ones, extremely
rare from point of view of classical statistics, obtain appreciable probability
and can be observed.  They were observed in the cosmic rays experiments
(Centauro events) at energy $\sim 1000$ TeV. This energy region will be covered
and studied in detail in the LHC experiments.

The consequences of the Bose-Einstein statistics for pion creation and their 
charge distributions is being discussed in the literature for a rather long
period of time (see, e.g., the paper  [{\it I}] in the list below). A
possibility of formation of a large domain of disoriented chiral condensate 
was indicated in the work [{\it II}] and then in  [{\it III,IV}]. The more
detailed lists of references devoted to this subject can be found in the more
recent publications, see, e.g., the papers [{\it V,VI,VII}].

As far as we aware, our paper [{\it a}] was one of the first  where the  pion
charge distributions were calculated, proceeding directly from the
Bose-Einstein statistics, and where the analogy with the induced photon
emission was strongly emphasized. However because of some technical reasons
this paper  was never sent to any physical journal. Its short version was
published in [{\it b}] (in Russian) and is practically inavailable. Therefore
we put now the paper [{\it a}] in the LANL archive. No changes have been done
in this text except for correcting some misprints.   \vspace{0.5cm}

\noindent
[{\it a}] V.A. Karmanov and A.E. Kudryavtsev,  {\it The Centauro events as a
result of induced pions emission}, Preprint ITEP-88,  Moscow, 1983.\\

\noindent
[{\it b}] V.A. Karmanov and A.E. Kudryavtsev, {\em The charge distributions of
creation of coherent pions}, In: ``Nucleon-nucleon and hadron-nucleus
interactions at intermediate energies", Proc. of third symp. of LINP,
Leningrad, 1986, p. 558 (in Russian). 

\newpage

\noindent
[{\it I}] D. Horn and R. Silver, Ann. of. Phys. (NY), {\bf 66} (1971) 509.\\

\noindent
[{\it II}] A.A. Anselm, Phys. Lett., {\bf B217} (1989) 169.\\

\noindent
[{\it III}] J.-P. Blaizot and A. Krzwicki, Phys. Rev., {\bf D46} (1992) 246.\\

\noindent
[{\it IV}]  J. Bjorken, K. Kowalski and C. Tailor, SLAC-PUB-6104; Les
Rencontres
de Physique de la Valle d'Aoste, {\it Results and Perspectives in Particle
Physics}, La Thuile, 1993, ed. M. Greco (Editions Frontieres, 1993).\\

\noindent
[{\it V}] S. Pratt and V. Zelevinsky, Phys. Rev. Lett., {\bf 72} (1994) 816.\\
 
\noindent
[{\it VI}] I.I. Kogan, JETP Lett., {\bf 59} (1994) 307.\\ 

\noindent
[{\it VII}] R. Amado, Yang Lu, Phys. Rev., {\bf D54} (1996) 7075.\\ 

\newpage
\begin{flushright}
ITEP - 88
\end{flushright}
\begin{center}
INSTITUTE\; OF\; THEORETICAL\\ AND\; EXPERIMENTAL\;  PHYSICS
\end{center}

\vspace{4cm}
\large
V.A. Karmanov$^*$, A.E. Kudryavtsev

\vspace{5cm}

\noindent
THE\; CENTAURO\; EVENTS\\
\\
AS\; A\;
RESULT\\
\\
OF\; INDUCED\; PIONS\; EMISSION
\normalsize

\vspace{6cm}

\begin{center}
 M O S C O W \;  1983\\
------------------------------------------------\\
$^*$Academy of Sciences of the USSR\\
P.N. Lebedev Physical Institute
\end{center}

\newpage


\begin{abstract}
It is shown that only in the case of $n$ pion production amplitude (at $n >
10$), which is symmetric enough relative to the momenta permutations, there is
appreciable probability of events with nothing but charged or nothing but 
neutral pions. Significant enhancement of the probability of these events in
comparison with that follows from Poisson distribution is caused by induced
pions emission, in complete analogy with the phenomenon of the induced photon
emission in QED. 
\end{abstract}

\section{Introduction}\label{intr}
Overwhelming majority of theoretical pictures describing the high energy
processes at high multiplicity leads to Poisson distribution for probability of
emission a given number of pions, provided the average number of pions is
fixed. In more sophisticated models this probability is given by the sum of
Poisson distributions - see papers \cite{aktm} and references therein. In all
the distributions of that kind the average number of neutral $n_0$ and charged
pions $n_{ch}$ are determined by the following relatons:
$$
n_0=\frac{1}{3}\bar{n},\quad n_{ch}=\frac{2}{3}\bar{n}.
$$
Notice that the experimanteal data obtained by the existing accelerators
($\sqrt{s}\leq 540$ GeV) do agree with that theoretical distributions. On the
other hand, in cosmic rays the  events were found with large multiplicity
($\bar{n} \sim 10 \div 100$), in which the neutral pions are absent
\cite{lfh,nik}. Their energy is about $E_{lab}\sim (1 \div 2)10^3$ TeV. These
events were called Centauros. Below we shall suppose that charged particles in
Centauro events are the pions.

Notice that it is practically impossible to interpret Centauro events as a
fluctuations of Poisson distribution - the probability  $w(\pi^+\pi^-)$ for
such a fluctuation is too small. For example, for $\bar{n}=100$ we get:
$$
w(\pi^+\pi^-)=e^{-n_0}\approx 10^{-14}.
$$

As it will be shown below, the charge distribution in a system of the given
total number of pions $n$ is closesly connected with symmetry properties of the
production  amplitude $M(p_1,\tau_1,\ldots,p_n,\tau_n)$ relative to
permutations of isotopic indices  $\tau_1,\ldots, \tau_n$. The later symmetry
leads also to certain type of symmetry for the permutations of momenta
$p_1,\ldots, p_n$, as it follows from Bose-properties for whole amplitude $M$.
Notice that the charge distribution differs from Poisson one if amplitude  $M$
is symmetric relative to permutation of momenta. As it will be seen below,  we
get in this case the significant increase of the probability of events with
neither neutral nor charged pions. On the other hand, the permutation
properties  of amplitude $M$ can give very useful information on properties of
the objects generating the Centauro events.

For a system of $n$ pions with small values of total isospin the charge
distributions for certain types of symmetries (Young tableaux) were derived
more that twenty years ago by Pais \cite{pais}\footnote{In ref. \cite{pais}
concrete realizations for all Young tableaux with one and two rows were
constructed, some special cases with three lines were considered also. In
connection with this question see also paper \cite{nikishov}.}. The charge
distribution in the case, when all possible Young tableaux give the same
contribution to the wave function, was obtained in \cite{cerulis}. In this case
the distribution has sharp form and is close to Poisson one for large values of
$n$.

Recently in connection with Centauro events in ref. \cite{andreev} the role of
small isospins in a system of $n$ pions was emphasized. This is an important
fact because in the nuclear collisions namely small vlues of isospin are
realized.  To explain Centauro events in ref. \cite{andreev} the most symmetric
form of wave function for a system of $n$ pions was also used. These two
assumptions automatically lead to the  smooth charge distribution, in
accordance with ref.  \cite{pais}.

In the present paper we are going to collect altogether those results of refs.
\cite{pais,cerulis,andreev}, which in our opinion may take immediate attitude
toward Centauro problem.  Transparent deduction of the results
\cite{pais,cerulis,andreev} will be given. Taken altogether, these results give
rigid restrictions on theoretical models of Centauros. We consider this paper
as a short review.

Physical reasons will be explained, which enhance the production  probability
of the pions of the same sort. This enhancement takes place when the totally
symmetric (in momentum space) amplitude dominates, and it compensates
Poisson's  diminition of process. {\it This phenomenon has much in common with
induced photon emission, that leads to formation of laser bunch.}

\section{Connection between the symmetry of amplitude and the charge
distribution of pions}\label{connect}

The production amplitude of $n$ pions has the form:
\begin{equation}\label{eq1}
M(p_1,\tau_1;p_2,\tau_2;\ldots;p_n,\tau_n)
=\sum_{\alpha}M_{\alpha}(p_1,\ldots,p_n)\varphi_{\alpha}(\tau_1,\ldots,\tau_n),
\end{equation}
where the sum is taken over all the possible types of symmetry $\alpha$,
characterized by Young tableaux. As we have only three sorts of pions
($\pi^+,\pi^0,\pi^-$), index $\tau_i$ ($i=1,2,...,n$) takes also only three
nonzero values. It follows that the rows number in Young tableaux  doesn't
exceed three as well. Totally symmetric function
$\varphi_{\alpha}(\tau_1,\ldots,\tau_n)$ corresponds to the only Young tableau
with one row.

The most important fact consists in the following: more symmetric function
$\varphi_{\alpha}(\tau_1,\ldots,\tau_n)$ gives more smooth distribution in
charge space. On the contrary, less symmetric Young tableaux with three rows
give sharp charge distribution. Notice that the total number of young tableaux
is very large:
$$ N\approx\frac{3}{8}\sqrt{\frac{3}{\pi}}\frac{3^n}{n^{3/2}} $$
 and most
part of $N$ comes from Young tableaux with three rows. So to observe the
Centauro events with appreciable probability, it is necessary  to increase the
role of symmetric terms (with one and two rows) in the sum (\ref{eq1}).  In
other words, it means that by some dynamical reasons the pions are produced
in symmetric states in momentum space.

First of all, let us demonstrate the pion distribution in the case when
large number of amplitudes with the different symmetry types gives the same
contribution into the sum (\ref{eq1}). Consider the case of even number $n=2k$
with total zero isospin $I=0$. We shall obtain the pions distribution for the
amplitude in a special form:
\begin{equation}\label{eq2}
M=\hat{S}\left\{M(p_1,p_2,\ldots,p_n)(\vec{a}_1\cd\vec{a}_2)\ldots
(\vec{a}_{2k-1}\cd\vec{a}_{2k})\right\},
\end{equation}
where $\vec{a}_i$ ($i=1,\ldots, n$) is isotopic vector of pion, $\hat{S}$ is
the symmetrization operator. Equation (\ref{eq2}) for $M$  is the special form
of (\ref{eq1}), as eq. (\ref{eq2}) is symmetric relative to  permutations of
pions in pairs, e.g., $\vec{a}_1\leftrightarrow \vec{a}_2$, etc. Nevertheless
this form of $M$ gives rather sharp distribution in number of neutral or
charged pions. Suppose that $M(p_1,p_2,\ldots,p_n)$ depends  sharply on the
arguments $p_1,p_2,\ldots,p_n$ so as it differs noticeably  from zero only when
$p_1\approx \bar{p}_1,p_2\approx \bar{p}_2,\ldots$ and
$\bar{p}_1\neq\bar{p}_2\neq\bar{p}_3\neq\ldots \bar{p}_n$. This condition
provides that the amplitude $M$, eq. (\ref{eq2}), contains a large number of
terms with different permutation symmetries. Making symmetrization, we get
$M(p_2,p_1,p_3,\ldots,p_n)\approx 0$ for  $p_1\approx \bar{p}_1$, $p_2\approx
\bar{p}_2$. On the contrary,  $M(p_2,p_1,p_3,\ldots,p_n)$ differs from zero
when  $p_1\approx \bar{p}_2$, $p_2\approx \bar{p}_1,$ $p_3\approx
\bar{p}_3,\ldots$. We get the same result making permutation of any pair of
arguments. So, integrating over whole phase volume to calculate the cross
section,  we obtain that all the permuted terms in eq. (\ref{eq2}) give the
same contribution. It follows that the value of cross section is proportional
to a number of nonzero isospin amplitudes in eq. (\ref{eq2}).

Now let us find out how this number depends on the number of charged (or
neutral) pions. Assume that among $n=2k$ pions there are $k_1$ of $\pi^+$,
$k_1$ of $\pi^-$ and $2k_2$ of $\pi^0$-mesons. Considering vectors  $\vec{a}_i$
in (\ref{eq2}) in the coordinates $a_{\pm},a_0$, we get for the isospin part of
amplitude:
\begin{equation}\label{eq3}
\chi=\left[\delta^{-\tau'_1}_{\tau_1}\cdots
\delta^{-\tau'_{k_1}}_{\tau_{k_1}}\right]
\left[\delta^{-\sigma'_1}_{\sigma_1}\cdots
\delta^{-\sigma'_{k_2}}_{\sigma_{k_2}}\right].
\end{equation}
As it is seen, $\chi=1$ for $\tau_1=\ldots=\tau_{k_1}=1$,
$\tau'_1=\ldots=\tau'_{k_1}=-1$, $\sigma_1=\ldots=\sigma_{k_2}=\ldots=
\sigma'_{k_2}=0$. Let's look for such  permutations of mesons, which give us
nonzero values of $\chi$. First we can do all the possible permutations of
upper indices in the first brackets in (\ref{eq3}), the set of lower indices
being fixed, and  obtain the factor $k_1!$. Secondly it is also possible to
permute indices in second bracket in eq. (\ref{eq3}). This operation changes
positions of $\pi^0$-mesons (no identical operations have been done!). The
number of nonzero amplitudes is $(2k_2-1)!!$. Permutations of indices between
both brackets give zero result.\footnote{Total number of terms in (\ref{eq2})
and in the cross section also is $2^k k! k_1 ! (2k_2-1)!!$. Later on we shall
study the distribution over $k_1$ and $k_2$ only.} From this it follows that
the probability $w(k_1,k_2)$ to emit $k_1$ pairs of $\pi^+\pi^-$ and $k_2$
pairs of $\pi^0$ is
\begin{equation}\label{eq4}
w(k_1,k_2)\sim \frac{k_1 !\; (2k_2-1)!!}{(k_1 !)^2\;(k_2 !)^2}.
\end{equation}
The denominator in (\ref{eq4}) appeared from the usual normalization  condition
for the cross section with an identical particles in final state, see, for
instance, ref. \cite{bdll}. Calculating the normalization constant for
$w(k_1,k_2)$, we finally get:
\begin{equation}\label{eq5}
w(k_1,k_2)=\frac{k!}{k_1 !\; k_2 !}\left(\frac{2}{3}\right)^{k_1}
\left(\frac{1}{3}\right)^{k_2},
\end{equation}
where $k=k_1+k_2$. This is binomial distribution.
For large values of $k$ and small $k_1$ (or $k_2$) we may approximate
eq. (\ref{eq5}) by Poisson distribution. It follows from eq. (\ref{eq5})
that the emission probability of nothing but charged particles only (i.e., the
probability to observe the Centauro event) is suppressed strongly, e.g.,
$w(\pi^+\pi^-)\approx 10^{-9}$ for $n=100$.

In the deduction given above we used for amplitude $M$ partially symmetric
form, eq. (\ref{eq2}). Surely there exist other combinations of isovectors
$\vec{a}_i$ of less symmetric form, e.g.  $\vec{a}_1\cd\left[\vec{a}_2\times
\vec{a}_3\right]$.

Being included in $M$, they would give more sharp charge distribution then it
follows from (\ref{eq5}). In the most consistent form tetrms of all different
symmetries were taken into consideration in ref. \cite{cerulis}. Contribution
to the cross section from all that terms was taken to be equal. In comparison
with (\ref{eq5}) the asymptotics of formula (16) from ref.  \cite{cerulis}
gives the probability of event without charged pions
\begin{subequations}
\begin{equation}\label{eq6a}
w(\pi^0)=\frac{8}{3}\sqrt{\frac{\pi}{3}}\frac{\sqrt{n}}{3^n}
\end{equation}
Analogously obtained probability of nothing but charged particles
has the form:
\begin{equation}\label{eq6b}
w(\pi^+\pi^-)=\frac{8}{3}\sqrt{\frac{2}{3}}\left(\frac{2}{3}\right)^n
\end{equation}
\end{subequations}
This is less than it follows from eq. (\ref{eq5}) at $n=2k$.

Let us consider now the contrary case. We find the charge distribution in a
system of $n=2k$ pions at $I=0$, described by the state totally symmetric in
the momenta (and isospin) variables. Corresponding isospin wave function is
obtained from eq. (\ref{eq2}) provided the amplitude $M(p_1,\ldots,p_n)$ is
not changed under the momenta permutations. We emphasize that in this case the
amplitude (\ref{eq2}) is not yet the amplitude of a particular form. In this
case it is defined unambiguously and doesn't depend on the angular momentum
addition scheme in the initial amplitude which produces after symmetrization
the amplitude eq. (\ref{eq2}). Therefore all nonzero terms in eq. (\ref{eq2})
are the same, that results in 100\% constructive interference. Then instead of
eq. (\ref{eq4}) we obtain
\begin{equation}\label{eq7}
w(k_1,k_2)\sim \frac{(k_1 !)^2\;[(2k_2-1)!]^2}{(k_1!)^2\;(2k_2)!}.
\end{equation}
Calculating the normalization factor by means of the relation:
$$
\sum_{l=0}^k\frac{(2l-1)!!}{(2l)!!}=\frac{(2k+1)!!}{(2k)!!},
$$
where, by definition, $(-1)!!=1$, we find:
\begin{equation}\label{eq8}
w(k_1,k_2)=\frac{2^k k!}{(2k+1)!!}\;\frac{(2k_2-1)!!}{2^{k_2} k_2!}.
\end{equation}

\begin{figure}[!ht]
\begin{center}
\mbox{\epsfysize=10.cm\epsfxsize=15.cm \epsffile{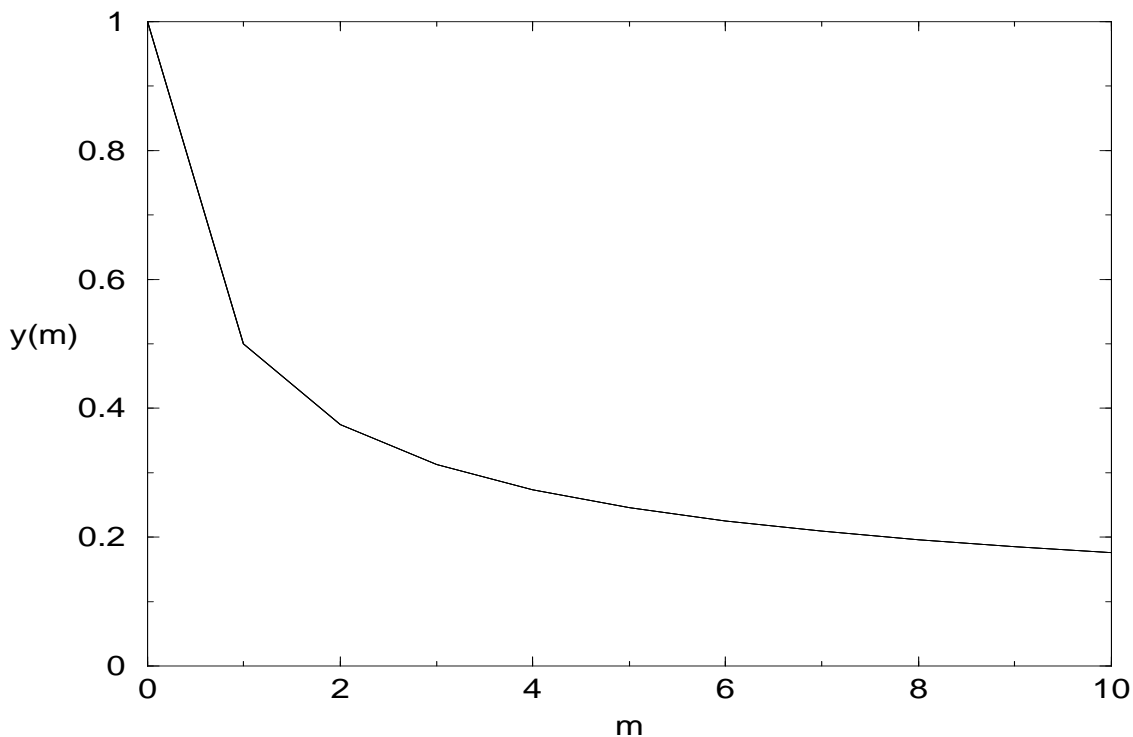}}
\caption{The function $y(m)=(2m-1)!!/(2^m m!)$. It gives multiplicity
distribution over a number of paires of neutral pions.\label{fig1}}
\end{center}
\end{figure}

In contrast to eq. (\ref{eq5}), the distribution (\ref{eq8}) is rather smooth.
It is shown in figure \ref{fig1}. At $k_2\gg 1$ we have $w(k_2)\sim
1/\sqrt{k_2}$. The average numbers of the neutral and charged pions both in the
case of distribution (\ref{eq8}) and in the case of eq. (\ref{eq5}) are
found the same and equal to  $$ n_0=\frac{1}{3}n,\quad n_{ch}=\frac{2}{3}n. $$
>From eq. (\ref{eq7}) at $n\gg 1$ we obtain the production probabilities of
nothing but neutral pions $w(\pi^0)$ or nothing but charged pions
$w(\pi^+\pi^-)$ (Centauros):
\begin{equation}\label{eq9}
w(\pi^0)=\frac{1}{n},\quad
w(\pi^+\pi^-)=\sqrt{\frac{\pi}{2n}}.
\end{equation}

These probabilities are enhanced by many orders in comparison to eq.
(\ref{eq5}). This enhancement is just well known in quantum optics the induced
emission phenomenon. The numerator of eq. (\ref{eq7}) contains extra (in
comparison with eq. (\ref{eq4})) factorials $k_1!\;(2k_2-1)!!$. These extra
factorials  appear due to interference, which is extremely essential in the
case of symmetric state relative to the final pion momenta permutations.   The
same behaviour of amplitude is found in the case of pion emission with the same
momenta. The reason of this enhancement results from the fact that the emission
of one sort bosons  in the state with the same momenta has essential advantage
over the emission of the same number of bosons of two or several sorts.  Note
that according to eq. (\ref{eq9}) the probability $w(\pi^0)$ by $(\pi
n/2)^{1/2}$ times less than $w(\pi^+\pi^-)$, in contrast to the Poisson
distribution. In the latter case we have:
$$
w(\pi^0)\ll w(\pi^+\pi^-).
$$
However, inspite of maximal enhancement of neutral pion production probability
$w(\pi^0)$, it  remains less than $w(\pi^+\pi^-)$. This fact is connected with
particular diminution of "initial" Poisson probability $w(\pi^0)$.

This can be also explained without starting from the Poisson distribution.
Although the enhancement of induced $\pi^0$-emission is considerably stronger
than that in the case of the same number of $\pi^+\pi^-$ (see below eq.
(\ref{eq10})), however the combinatorial factor (the coefficient 2) enhances
the  $\pi^+\pi^-$-production (see below eq. (\ref{eq11})). This factor appears
in expansion of the scalar operator  $\vec{c}^{\;\dagger
2}=2c^{\dagger}_{\pi^+} c^{\dagger}_{\pi^-}+c^{\dagger 2}_{\pi^0}$ in terms of
the pion creation operators.

In order to emphasize more clearly the connection between possible Centauro
production and the induced photon emission we reproduce eq. (\ref{eq8}) by
means of  the second quantization formalism. The totally symmetric isospin
$n$-particle wave function at $I=0$ has the form:
$$
\vert\phi\rangle = \frac{1}{\sqrt{(2k+1)!}}\left(\vec{c}^{\;\dagger 2}\right)^k
\vert0\rangle,
$$
where $\vec{c}^{\;\dagger 2}=2c^{\dagger}_{\pi^+} c^{\dagger}_{\pi^-}
+c^{\dagger 2}_{\pi^0}$, $c^{\dagger}_i$ is the pion creation operator of
$i$-th  sort, $n=2k$. Calculating the amplitude
$\langle k_1\pi^+,k_1\pi^-,2k_2\pi^0\vert\phi\rangle$ and taking into account
that
\begin{equation}\label{eq10}
\langle k_1\pi^+\vert (c^{\dagger}_{\pi^+})^{k_1}\vert 0 \rangle=
\sqrt{k_1 !}\ ,
\quad
\langle 2k_2\pi^0\vert (c^{\dagger}_{\pi^0})^{2k_2}\vert 0 \rangle=
\sqrt{(2k_2) !}\ ,
\end{equation}
we find:
\begin{equation}\label{eq11}
\langle k_1\pi^+,k_1\pi^-,2k_2\pi^0\vert\phi\rangle
=C^{k_1}_k\frac{2^{k_1}k_1!\;\sqrt{(2k_2)!}}{\sqrt{(2k+1)!}},
\end{equation}
where $C^{k_1}_k$ is the binomial coefficient. The amplitude (\ref{eq11})
squared reproduces eq. (\ref{eq8}). Enhancement of the pion production
amplitude in the case of the same sort of pions is due to the same factors
(\ref{eq10}) which enhance the photon induced emission amplitude.

Note that the formulae (\ref{eq9}) for $w(\pi^0)$ and $w(\pi^+\pi^-)$ coincide
with the result obtained in ref. \cite{andreev}\footnote{In ref.
\cite{andreev} the probabilities $w(\pi^0)$ and $w(\pi^+\pi^-)$ were calculated
by projecting pion field on the state of small isospin, $I=0,1$.} with accuracy
of replacement of the pion number $n$ in eq. (\ref{eq9}) by the average number
$\bar{n}$ in ref. \cite{andreev}. The probability of a given number of pions in
ref.  \cite{andreev} was described by the Poisson distribution. Averaging eqs.
(\ref{eq9}) by Poisson distribution just leads to the replacement of $n$ by
$\bar{n}$ (with accuracy of the power corrections).

\section{Role of the small total isospin of the $n$ pion system}

It was shown in the previous section that refusal from domination of most
symmetric isospin wave function  leads to negligibly small probability of
Centauro production. In the present section we show that refusal from  small
isospin values also sharply diminishes the probabilities $w(\pi^0)$ and
$w(\pi^+\pi^-)$ even in the case of totally symmetric wave function. This
aspect of phenomenon has been emphasized in ref. \cite{andreev}.

Les us calculate the probabilities $w(\pi^0)$ and $w(\pi^+\pi^-)$ considering
the totally symmetric $n$-pion state with $n=2k$ and isospin
$I=n,I_3=0$.\footnote{The charge structure of the state with $I_3$  close to
the maximal value is almost definite. Therefore the problem appears for the
state with small $I_3$ only.} The isospin wave function  has the form:
\begin{equation}\label{eq12}
\vert \phi_I\rangle =A(I)(\vec{c}^{\;\dagger2})^k \;Y_{I0}
\left(\frac{\vec{c}^{\;\dagger}}{\sqrt{\vec{c}^{\;\dagger 2}}}\right)
\vert 0 \rangle,
\end{equation}
where $A(I)$ is the normalization factor:
$$
A(I)=\sqrt{\frac{4\pi}{(2I+1)!!}}.
$$
The probabilities concerned are determined by the following amplitudes:
\begin{eqnarray*}
M(\pi^0)&=&\langle 2k\pi^0\vert \phi_I\rangle =
A(I)\sqrt{\frac{2I+1}{4\pi}}P_I(1)\;\langle 2k\pi^0\vert
(c^{\dagger}_{\pi^0})^{2k}\vert  0\rangle,
\\
M(\pi^+\pi^-)&=&\langle k\pi^+,k\pi^- \vert \phi_I\rangle =
A(I)\sqrt{\frac{2I+1}{4\pi}}P_I(0)\; 2^k
\langle k\pi^+,k\pi^- \vert
(c^{\dagger}_{\pi^+}c^{\dagger}_{\pi^-})^k\vert  0\rangle,
\end{eqnarray*}
where $P_I(x)$ is the Legendre polynom.

>From here at $I\gg 1$ we obtain:
\begin{subequations}
\begin{eqnarray}
w(\pi^0)&=& \frac{\sqrt{\pi I}}{2^I},
\label{eq13a}
\\
w(\pi^+\pi^-)&=&\frac{\sqrt{2}}{2^I}.
\label{eq13b}
\end{eqnarray}
\end{subequations}
At $I=n\approx 100$ the probabilities (\ref{eq13a}), (\ref{eq13b})
have the order of $10^{-30}$.

\section{Conclusion}\label{concl}

Thus we have convinced that the events containing nothing but
$\pi^+\pi^-$-mesons or $\pi^0$-mesons at high multiplicity can have appreciable
probability only at simultaneous realization of the following two conditions:
\begin{enumerate}
\item Symmetric state of $n$-meson system relative to the permutation of
momenta (and isospin indices).
\item
Small total isospin.
\end{enumerate}

The first condition seems us to be considerably more essential than the second
one, since due to isospin conservation the second condition is satisfied
automatically, while the first condition implies the strong restrictions to the
dynamics of high multiplicity pion production.  We emphasize that these results
follow from rather general consideration.  Therefore any model, which pretends
to explanation of Centauros,  must first of all satisfy to two these
conditions.\footnote{We mean the models in which the hadrons produced are
considered as pions.}

It should mention that under symmetric function we mean the function
corresponding to the Young tableau with one or two rows. Thus, proceeding from
ref. \cite{pais} one can obtain that $I=0$ for the Young tableau with two equal
rows $w(\pi^+\pi^-)=1/n$, but $w(\pi^0)=0$. However, the Young tableau with
three approximately equal rows leads to distribution which differs from zero
only at $n_0$ and $n_{ch}$ close to $\frac{1}{3}n$ and $\frac{2}{3}n$
correspondingly. Since the overwhelming majority of the Young tableaux  is the
Young tableaux with three rows, namely they determine the distribution when all
the Young tableaux give equal contributions.

On these grounds it can be conceived that Centauros are produced from decay of
a particle as if consisting from large number of pions in totally symmetric
state. On the ground that  the experimental hadron transverse momenta in
Centauro equal to $p_T\geq 1.5$ GeV/c, one can expect that the size of this
hypothetical object is of the order of (1.5 GeV/c)$^{-1}$. The field with
energy $E\approx 230$ GeV \cite{nik} is concentrated in this volume.
Considering the pions as quanta of this field one can suppose that this object
could be similar in its properties to the classical solution of a field
equation.  One can expect that namely at high energy the heavy particles having
the properties close to the classical solutions can be produced.  This
hypothesis  have been suggested in ref. \cite{shapiro}. Note that possible
soliton solution  to explain Centauros is, for example, the slowly  damping
classical solution for Higgs scalar field found in ref. \cite{sasha}.

Let us give a possible estimation of the Centauro production cross section.
Assuming that a Centauro is produced from a soliton decay, we see that the only
dimensional parameter is its size $r_0\sim \mbox{(1.5 GeV/c)}^{-1}$. It follows
that relative probability of the soliton production is the ratio:
\begin{equation}\label{eq14}
w\sim r_0^2/r_{st}^2\sim 10^{-2},
\end{equation}
where $r_{st}$ is the strong interaction radius. This estimation is rather
rough. The probability of Centauro production is obtained by multiplying eq.
(\ref{eq14}) by (\ref{eq9}). Therefore the probability to find a Centauro is
estimated as
\begin{equation}\label{eq15}
w(\mbox{Centauro})\sim 10^{-3}.
\end{equation}

Although this estimation does not depend explicitly on energy, it becomes valid
only begining from rather high energy (which we don't estimate), when
quasiclassical objects can be produced.  CERN SPS-collider experiments
\cite{alpgard, arnison} at $\sqrt{s}=540$ GeV give so far no evidences for
Centauro production. From the present point of view this means that inspite the
fact that the energy  $\sqrt{s}=540$ GeV exceeds the threshold for production
of the mass $M=230$ GeV/c$^2$, this energy is not enough for proceeding the
physical phenomena leading to Centauro production (e.g., for developing the
soliton object).

The other approaches to Centauros were discussed in refs. \cite{nik,fowler}.

\section{Appendix}\label{app}

From the present consideration some conclusions follow for the high
multiplicity pion production in nuclear reactions. It follows that charge
distribution of pions produced in any nuclear reaction would be smooth provided
the pion momenta are close enough to each other. The latter comes true in
kinematics near the phase volume bound. For example, in the reaction
\begin{equation}\label{eq16}
pp\rightarrow pp+n\pi
\end{equation}
at highest possible pion number $n$ allowed by energy conservation the reaction
kinematics automatically forbids the wide variation  of the pion momenta. This
ensures automatically the total symmetry of pion production amplitude relative
to the momenta permutation in the narrow allowed region. As it was shown
above,  we obtain the smooth charge distribution in this case.

It would be useful to carry out the experimental research of the reaction
(\ref{eq16}) in the kinematics concerned. Such investigation  is possible at
the meson factory accelerators, where due to high intensity the small cross
sections can be measured.

Another interesting reaction is the extremely many pion annihilation of low
energy antiprotons (see ref. \cite{dalkarov}):
\begin{equation}\label{eq17}
p+\bar{p}\rightarrow n\pi.
\end{equation}

Usually rather sharp binomial distribution (as eq. (\ref{eq5})) is expected in
such processes, which follows from the statistical models \cite{muirheid}. In
this reaction at highest possible multiplicity $n$ we expect a smooth
distribution as well.

The authors are grateful to A.B. Kaidalov, V.E. Markushin, A.I. Nikishov, V.A.
Novikov, L.B. Okun, I.S. Shapiro and M.B. Voloshin for useful discussions and
valuable remarks.


\end{document}